\documentclass[12pt,aps,pre,superscriptaddress]{revtex4}

\usepackage{amsmath, amsthm, amssymb}
\usepackage{graphicx}

\usepackage{times}

\begin{document}
\title{Gradual Regime Shifts in Fairy Circles}

\author{Yuval R. Zelnik}
\affiliation{Department of Solar Energy and Environmental Physics, BIDR, Ben-Gurion University of the Negev, Sede Boqer Campus 8499000, Israel}
\author{Ehud Meron}
\affiliation{Department of Solar Energy and Environmental Physics, BIDR, Ben-Gurion University of the Negev, Sede Boqer Campus 8499000, Israel}
\affiliation{Department of Physics, Ben-Gurion University, Beer
Sheva, 84105, Israel}
\author{Golan Bel}
\email{bel@bgu.ac.il}
\affiliation{Department of Solar Energy and Environmental Physics, BIDR, Ben-Gurion University of the Negev, Sede Boqer Campus 8499000, Israel}

\begin{abstract}
Large responses of ecosystems to small changes in the conditions--regime shifts--are of great interest and importance. In spatially extended ecosystems, these shifts may be local or global. Using empirical data and mathematical modeling, we investigated the dynamics of the Namibian fairy circle ecosystem as a case study of regime shifts in a pattern-forming ecosystem. Our results provide new support, based on the dynamics of the ecosystem, for the view of fairy circles as a self-organization phenomenon driven by water-vegetation interactions. The study further suggests that fairy circle birth and death processes correspond to spatially confined transitions between alternative stable states. Cascades of such transitions, possible in various pattern-forming systems, result in gradual rather than abrupt regime shifts.
\end{abstract}
\maketitle

The response of ecosystems to climate variability and anthropogenic disturbances is a fundamental aspect of ecology.
Much attention has been devoted recently to large responses of ecosystems to small environmental changes or disturbances.
Such responses, often termed ``catastrophic regime shifts,'' are conceived of as abrupt transitions between two alternative stable states that occur uniformly across the ecosystem\cite{Scheffer2001nature}.
Spatially extended systems can respond in different ways to varying conditions, including pattern formation\cite{rietkerk2004science,Deblauwe2008geb,Meron2012ecomod} and spatially confined transitions to alternative stable states\cite{Bel2012theo_ecol}.
When one of the alternative stable states is spatially patterned, a multitude of additional stable states can appear, each a mosaic of fixed domains that alternate between the uniform and the patterned states \cite{Lejeune2002pre,Knobloch2008nonlinearity,Meron2012ecomod}. 
The existence of such hybrid states can strongly affect the dynamics of state transitions in fluctuating environments\cite{Gandhi2015prl}, and may lead to gradual rather than abrupt shifts\cite{Bel2012theo_ecol,zelnik2013regime}.
This prediction of pattern formation theory has never been tested either in a real ecosystem or in a model describing the dynamics of a specific ecosystem.

Pattern formation is widespread in natural ecosystems\cite{rietkerk2008tree,Deblauwe2008geb}, but good case models for studying gradual regime shifts are not abundant. 
An outstanding candidate is the Namibian fairy circle (NFC) ecosystem, which is fairly homogeneous and undisturbed, and recently has been the subject of intense research \cite{tlidi2008lnp,picker2012ants,tschinkel2012life,juergens2013biological,cramer2013namibian,Fernandez-Oto2014philos_trans_A,getzin2015adopting}. 
The NFC ecosystem consists of a uniform matrix of perennial grass, punctured by circular gaps of sandy bare soil--the fairy circles--that on landscape scales, form nearly periodic patterns\cite{picker2012ants,getzin2015adopting}.

Various explanations for the formation of fairy circles have been suggested, including the release of poisonous gas and the feeding habits of ants and termites\cite{becker2000fairy,juergens2013biological,getzin2015adopting}.
These explanations, however, have not uncovered, as of yet, the small-scale feedbacks needed to account for the emergence of the large-scale order\cite{rietkerk2004science,rietkerk2008tree,Meron2012ecomod}.
On the other hand, it is well established by model studies and confirmed by empirical observations that patch-scale biomass-water feedbacks can lead to regular landscape-scale vegetation patterns, and that periodic gap patterns, which highly resemble fairy circle patterns, 
can appear from uniform vegetation in response to water stress\cite{Hardenberg2001prl,Rietkerk2002an,Gilad2004prl,Borgogno2009gr,Deblauwe2008geb}.
Indeed, recent detailed comparisons of fairy circle remote sensing data and model gap patterns show high similarity in static properties, such as hexagonal symmetry and spatial correlations of fairy circle size and distance\cite{getzin2015adopting}.

In this paper, we combine a theoretical analysis of a vegetation model, fitted to the biotic and abiotic conditions of the NFC ecosystem, with an empirical data analysis, to account for fairy circle dynamics in the NamibRand Nature Reserve from the years 2004 to 2013, thereby accomplishing two goals. 
The first goal is to substantiate the view of fairy circles as a pattern-formation phenomenon by complementing the current statistical evidence\cite{tlidi2008lnp,cramer2013namibian,getzin2015adopting} with evidence based on the dynamics of the NFC ecosystem.
The second goal is to demonstrate the feasibility of gradual regime shifts in the NFC ecosystem as cascades of unidirectional transitions across hybrid states.

\section{Model}
The model we use is based on the Gilad et al. vegetation model\cite{Gilad2004prl, Gilad2007jtb}, which captures three different mechanisms of vegetation pattern formation\cite{Kinast2014prl}.
However, applying the model to the NFC ecosystem (sandy soil, confined root zones) results in a simplified model describing the dynamics of aboveground biomass ($B$) and soil-water ($W$) areal densities.
The simplified model captures a single pattern-forming feedback associated with the high rate of water uptake by the perennial grasses and the fast soil-water diffusion (relative to biomass expansion) in sandy soils.
This mechanism leads to higher soil-water content in the bare fairy circles as compared with the vegetation matrix\cite{Kinast2014prl}, in agreement with reported observations\cite{juergens2013biological,cramer2013namibian}.
The model equations read:
\begin{small}
\begin{eqnarray}
\label{eq:model}
\partial_T B &=& \Lambda WB(1 - B/K)(1 + E B)^2 - M B + D_B\nabla^2B \,, \label{SGeqb} \\
\partial_T W &=& P - N(1-R B/K)W - \Gamma WB(1 + E B)^2  + D_W\nabla^2 W\,. \nonumber \label{SGeqw}
\end{eqnarray}
\end{small}
In the biomass equation, $\Lambda$ is the biomass growth rate coefficient, $K$ is the maximal standing biomass,
$E$ is a measure for the root-to-shoot ratio, $M$ is the mortality rate, and $D_B$ represents the seed-dispersal or clonal growth rate.
In the soil-water equation, $P$ is the precipitation rate, $N$ is the evaporation rate, $R$ is a dimensionless factor representing a reduction of the evaporation rate due to shading,
$\Gamma$ is the water-uptake rate coefficient, and $D_W$ is the effective soil-water diffusivity in the lateral ($X,Y$) directions.
The model parameters used in this study are different than those used by Getzin et al. \cite{getzin2015adopting}, and were estimated from published data.
We refer the reader to the SI Appendix for more details on the derivation of the model and to the Methods and Materials section for details on the parameter estimations and values.

\section{Results}
We first study the stationary solutions, in one spatial dimension (1d), along the precipitation axis. Fig.1A shows uniform solutions representing bare soil and uniform vegetation and periodic solutions representing gap patterns. 
Within the bistability range of uniform vegetation and gap patterns, there is a subrange with many stable hybrid states, consisting of confined domains of uniform vegetation in an otherwise periodic vegetation pattern (and vice versa), (inset B).
The figure further shows 1d spatial profiles of a periodic gap pattern (panel C) and of a hybrid state that describes a periodic pattern with one missing gap (panel D). 
The corresponding 2d patterns are shown in panels E and F, and similar fairy circle patterns (obtained from satellite images) are shown in panels G and H.
Similarly to other models of dryland vegetation \cite{Borgogno2009gr,zelnik2013regime}, the model also exhibits spotted and striped patterns (see the SI Appendix), for lower values of precipitation.
\begin{figure}[h]
 \includegraphics[width=0.95\linewidth]{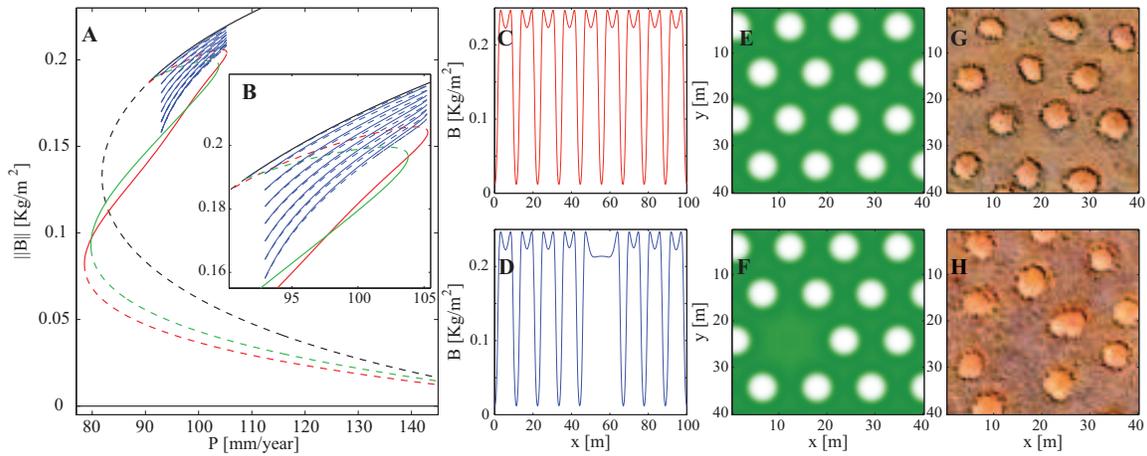}
 \caption{
 Steady states of the model. \textbf{A:} Partial bifurcation diagram for a 1D system, showing the $L^2$ norm of the biomass versus the precipitation rate.
The steady states shown include two uniform solutions of bare soil and uniform vegetation (black),
two periodic solutions of gap patterns with different wavelengths (red and green) spanning the bistability range of uniform vegetation and patterned states,
and hybrid solutions representing uniform vegetation with an increasing (as the branch snakes down) number of gaps (blue).
The solid (dashed) lines in the diagrams represent stable (unstable) solutions. \textbf{B:} Blowup of the hybrid-state range. \textbf{C:} The periodic solution corresponding to the red branch in \textbf{A}.
\textbf{D:} A hybrid solution corresponding to the lowest blue branch in \textbf{B} (a single missing gap) obtained with $P=102 [mm/yr]$.
\textbf{E,F:} The 2D versions of the solutions shown in \textbf{C,D}.
 \textbf{G,H:} Corresponding fairy circle patterns, obtained from the 2013 satellite image.
 }
 \end{figure}

Next, we use the model, along with high-resolution satellite images of the NamibRand area from 2004 to 2013, to suggest that fairy circle ``birth'' and ``death'' processes (appearance and disappearance of gaps)\cite{tschinkel2012life} correspond to transitions between hybrid states.
Figure 2A shows the birth of a fairy circle following a drought in 2007, while Fig. 2C shows the death of a fairy circle following a spate (a period of excess precipitation) in 2008.
Figure 2B and 2D show the same behaviors in model simulations, starting with initial conditions that mimic the initial fairy circle patterns in 2004 and applying precipitation downshift and upshift to mimic the drought and spate. 
The precipitation shifts were chosen to be strong enough to take the system out of the subrange of hybrid states (in opposite directions), thus inducing hybrid-state transitions.
\begin{figure}[ht]
 \includegraphics[width=0.95\linewidth]{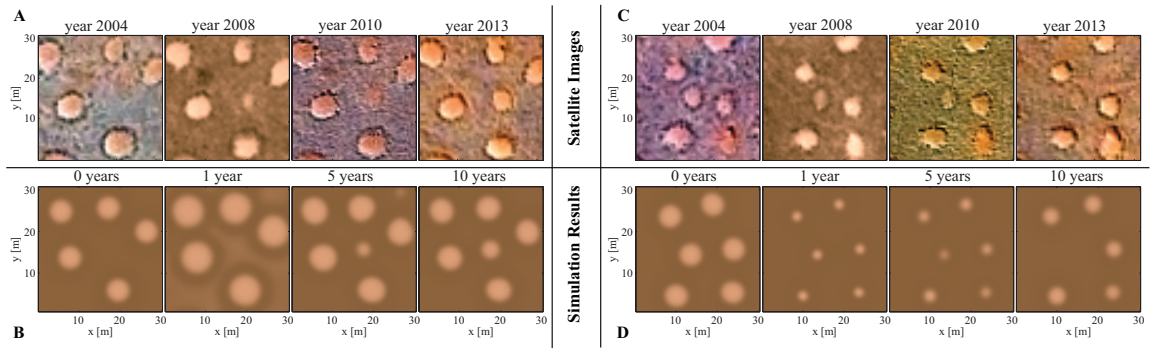}
 \caption{
  Birth and death of fairy circles as hybrid-state transitions.
 \textbf{A,C:} satellite images showing the dynamics of the appearance/disappearance of a bare soil domain (fairy circle birth/death).
 \textbf{B,D:} snapshots of corresponding simulations using initial conditions derived from the 2004 satellite images shown in panels \textbf{A} and \textbf{C}.
 Panels \textbf{B,D} show the simulated response to one year of drought/spate ($P=84 [mm/yr]$/$P=143 [mm/yr]$), after which the precipitation was set back to normal ($P=102 [mm/yr]$).
 The time shown at the top of the satellite image (simulation) denotes the year it was captured (number of years elapsed from the beginning of the drought or the spate).
 The time intervals between the simulation snapshots were chosen to emphasize the short-time-scale changes in gap size.
  All panels show the dynamics of a region of $30[m]\times30[m]$.
 }
\end{figure}

In addition to the appearance and disappearance of gaps, Fig. 2 also shows changes in gap sizes; a moderate drought expands the gaps, while a spate contracts the gaps. 
To gain a deeper understanding of these dynamics, we correlated fairy circle data that were extracted from satellite images with rainfall data from several meteorological stations in the NamibRand region, 
and compared the observed correlations with those predicted by the model. Figure 3 shows the cross-correlations between the accumulated rainfall over different time periods (integration time) and the average fairy circle size (panel A) and the fairy circle number (panel B). 
Analogous cross-correlations using model simulations (panels C and D) show a qualitative agreement. The simulation results are based on averaging over 100 regions and long time series, and therefore, they show smooth curves. 
Panels E and F show the same quantities for the model simulations but averaged over fewer regions and using shorter time series. 
The curves in these panels resemble the curves for the real data (panels A and B). The stronger correlations in the simulations (compared with those in the field) 
between the gap size and the precipitation is due to the fact that the simulated system is homogeneous, while the field is heterogeneous and also because the simulated precipitation is not intermittent as it is in the field.   
The results highlight the different time scales for changes in the size and in the number of fairy circles. For fairy circle size, the correlation magnitude is maximal for a relatively short integration time (of a few years), 
while for fairy circle number, the correlation magnitude does not decrease with time throughout the time range of available data. This suggests that the timescale for changes in the fairy circle size is much shorter than the time scale for changes in the fairy circle number.
\begin{figure}[ht]
 \includegraphics[width=0.52\linewidth]{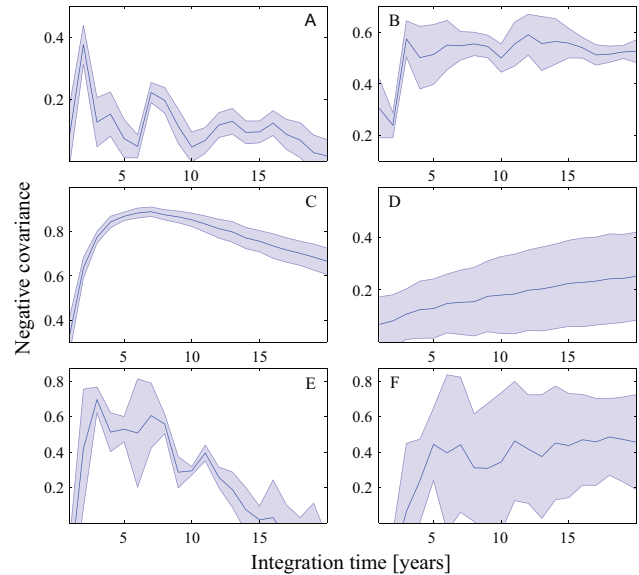}
 \caption{
 Correlations between the number and average size of fairy circles and the accumulated precipitation.
All graphs show the negative covariance versus the integration time for the accumulated precipitation (see the Materials and Methods section for precise definitions).
 \textbf{A,C:} correlations between average fairy circle size and accumulated rain for real data and simulation results, respectively.
 \textbf{B,D:} same information for the fairy circle number.
 Real data analyses (\textbf{A,B}) are based on eight satellite images (i.e., eight time points) and averaging over four regions.
 Simulation results (\textbf{C,D}) are based on 100 time points and averaging over 120 regions. Therefore, the curves for the simulation results are smoother.
 \textbf{E,F:} correlations for simulation data, similar to panels \textbf{C,D}, but including the same number of time points and regions as the real data.
 The shaded regions show the error assessment, calculated from the standard deviation of the sample.
  }
\end{figure}

We turn now to the question of gradual regime shifts and ask whether climate variability in the form of repeated short droughts can induce a gradual regime shift.
To this end, we simulated the model equations, starting with a uniform vegetation state that is locally disturbed to form a few gaps, and using precipitation downshifts that take the system outside the hybrid-state subrange periodically in time.
As Fig. 4A shows, such a precipitation regime drives the system towards a nearly periodic patterned state (rightmost panel) by successive shifts to hybrid states of larger gap numbers.
The opposite transition is shown in Fig. 4B, in which a series of short spates drives the system from a nearly periodic gap pattern, through a multitude of hybrid states, to a uniform vegetation state.
Systems initially in a uniform vegetation state (see the SI Appendix) or a perfectly periodic patterned state (Fig. 4C), exposed to the same series of droughts or spates, show only minor fluctuations in the biomass density and remain practically unchanged.
\begin{figure}[ht]
 \includegraphics[width=0.8\linewidth]{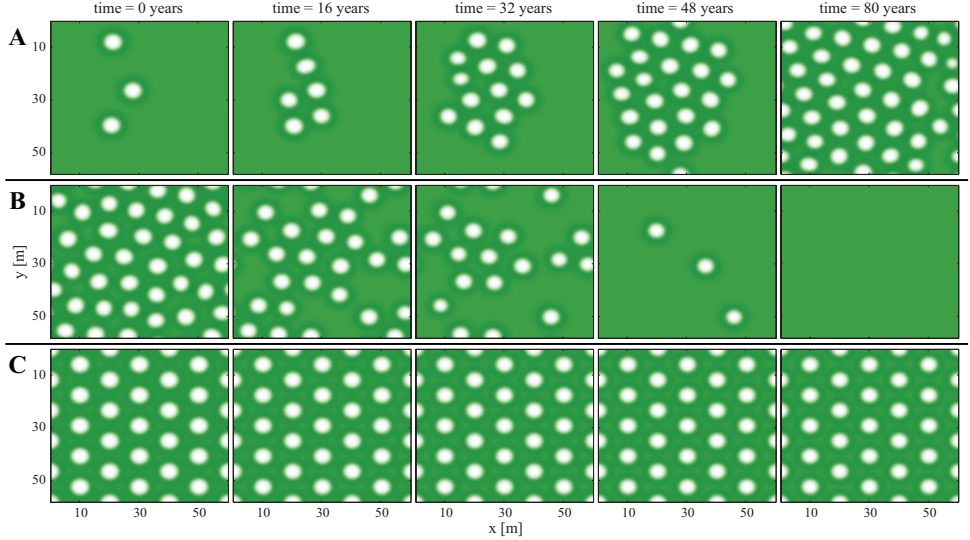}
 \caption{
 The response of 2D patterns to periodic short droughts or spates.
 \textbf{A:} a gradual regime shift from a stable hybrid state (leftmost panel) to a nearly periodic gap pattern (rightmost panel).
 The shift is driven by periodic one-year droughts with $P=81 [mm/yr]$, followed by 15 years of $P=102 [mm/yr]$.
 \textbf{B:} the reversed shift, namely a gradual shift from a nearly periodic pattern to uniform vegetation, driven by one-year spates with $P=138 [mm/yr]$ followed by 15 years with $P=102 [mm/yr]$.
The droughts/spates take the system out of the hybrid-state range (see Fig. 1A,B) and induce hybrid-state transitions in the course of which gaps appear/disappear, leading to a gradual regime shift.
 \textbf{C:} when the initial pattern is strictly periodic, no regime shift takes place even though the system is subjected to the same rainfall regime as in \textbf{B}.
 The simulated domains are $60 [m]\times56 [m]$.
 }
\end{figure}

\section{Discussion}
We regard the model results described above and their agreement with field observations as compelling evidence for the view of fairy circles as a pattern-formation phenomenon and for the existence of hybrid states in the NFC ecosystem.
This evidence is based on the following grounds:
\hfill\break\noindent
\emph{Spatial properties} -- the model, adjusted to the particular context of the NFC ecosystem, predicts gap patterns that resemble the spatial distribution of fairy circles in the field, as Fig. 1 shows.
Moreover, the model predictions agree with recent field observations of higher soil-water content within the fairy circles, relative to vegetated domains\cite{juergens2013biological,cramer2013namibian}.
\hfill\break\noindent
\emph{Dynamics} -- fairy-circle birth and death processes can be reproduced with model simulations by temporal escapes outside the existence range of hybrid states (Fig. 2), supporting the interpretation of these processes as hybrid-state transitions.
\hfill\break\noindent
\emph{Correlations with precipitation} -- significant correlations between the rainfall and the number and size of fairy circles indicate the importance of biomass-water interactions in the dynamics of the NFC ecosystem.
These correlations and the time scales associated with them, short for size variations and long for number variations, are in agreement with the model predictions (Fig. 3).
The time-scale difference can be understood using pattern formation theory; gap-size changes are related to the convergence to a steady state, while gap-number changes are related to the dynamics of fronts that separate domains of uniform vegetation and gap patterns\cite{Gandhi2015prl}, structural and wavelength changes of the patterned state, and other nonlinear processes.

This evidence does not rule out alternative hypotheses\cite{Vlieghe2015ecol_ento,juergens2015ecography}, but these hypotheses cannot be considered as satisfactory explanations of the fairy circle phenomenon unless they uncover the local feedbacks that are responsible for the emergence of large-scale order and account for the correlation of fairy circle dynamics with rainfall variability.
The model suggested here requires additional evidence in order to fully uncover the origin and dynamics of the NFC. First, measurements of the soil-water diffusivity and the dependence of the soil-water extraction rate on the vegetation density are required in order to support the mechanism suggested in our model. Second, the precipitation fluctuations in the field are very large and, according to the simplified model, are expected to drive the system towards one of the uniform states. We believe that accounting for the spatial heterogeneity of the ecosystem, i. e., topography, soil heterogeneity and other factors, would result in more robust patterns in the model. Third, the exact biological origin of hybrid states in the NFC system remains unclear. Controlled field experiments of the FCs dynamics, combined with the mathematical framework suggested here, will be able to provide conclusive evidence for the hypothesis proposed here and reveal the relevant biological mechanisms. In particular, experiments in areas with different soil textures, topography, and plant species, together with manipulations of the precipitation, are expected to better clarify the relative importance of specific plant physiology and ecosystem characteristics in shaping the observed NFC dynamics.

The results presented here are not limited to the NFC ecosystem and may be observed in other pattern-forming systems. The appearance of hybrid states in systems exhibiting a bistability of uniform and patterned states implies that transitions, from uniform vegetation to a gap pattern and vice versa, may take place through successive hybrid-state transitions driven by periodic escapes outside the existence range of hybrid states. Dynamics of this type allow for spatially extended ecosystems to exhibit gradual rather than abrupt regime shifts.

\section{Model Analysis}
 Vegetation pattern formation in this simplified model results from a finite wavenumber (Turing) instability of the uniform vegetation state. This instability requires strong water uptake, quantified by the parameters $\Gamma$ and $E$, and fast soil-water transport, characterized by $D_W$, relative to biomass dispersal, quantified by $D_B$\cite{Kinast2014prl}.
The high hydraulic conductivity, typical of sandy soils, acts in favor of this condition, but field observations point towards the possible role of an additional factor--the large underground termite populations found in the NFC ecosystem\cite{juergens2013biological}.
The underground network of channels that the termites form may increase soil-moisture transport in the lateral directions, leading to higher effective values of $D_W$.
Studying this possible termite effect is not only significant for understanding fairy circle formation, but also for understanding regime shifts, as the presence of termites may affect the existence and stability ranges
of the uniform and patterned vegetation states, as well as those of the hybrid states.

The bifurcation diagram in Fig. 1 A,B was calculated using a numerical continuation method (AUTO software).
The dynamics of the system was simulated using the pseudo-spectral method with periodic boundary conditions in Matlab, in either one or two dimensions.
The simulations in Fig. 2 were made using initial conditions of steady states of the system, each derived from a region of the 2004 satellite image.
This was done by taking the relevant region in the original image, segmenting it into vegetation and bare soil areas, and setting the values of the biomass and soil-water areal densities to those corresponding to the uniform solutions (uniform vegetation and bare soil, respectively).
The system was then integrated forward in time, with a constant value of precipitation rate, $P$, until a steady state was reached.

 \section{Parameter Estimations} 
 In all model simulations, we used the following parameter values:
$E=7 [m^2/kg]$, $K=0.4 [kg/m^2]$, $M=10.5 [1/yr]$, $N=15 [1/yr]$, $\Lambda=0.9 [(m^2/kg)/yr]$, $\Gamma=12 [(m^2/kg)/yr]$, $R=0.7$, $D_B=1.2 [m^2/yr]$, $D_W=150 [m^2/yr]$.
These values were either taken from data published in the literature or estimated using such data, as described below. The precipitation rate, $P$, was varied within the range $50<P<150 [mm/yr]$, which is similar to the range of annual rainfall in the NFC ecosystem.

The maximum standing biomass, $K$, was estimated based on maximal biomass densities reported for semi-arid savannah grasses\cite{nepolo2012short}.
The mortality rate, $M$, was estimated based on the mortality data of \emph{Stipagrostis Uniplumis} as a representative example of an abundant perennial grass in the NFC ecosystem\cite{zimmermann2010grass}.
The reported mortality rate was measured under normal conditions (i.e., including all the growth factors). The mortality rate in the model represents the natural mortality alone, and therefore, we used a higher mortality rate.
The biomass diffusion parameter, $D_B$, was estimated based on published clonal growth and seed dispersal data\cite{cain1997clonal}.
The value for the shading factor, $R$, was inferred from data on reduced evaporation by tree canopies\cite{wallace1999modelling}.
No relevant literature was found for the root-to-shoot parameter, $E$, for the perennial grass; the chosen value is similar to that used in an earlier study\cite{Gilad2007jtb} for woody dryland species.
The evaporation rate, $N$, was estimated using the relation $N=P/W$ for a uniform bare soil state, where the soil-water density, $W$, was determined from water-content sampling data inside a fairy circle\cite{juergens2013biological}.
The water-uptake rate coefficient, $\Gamma$, was inferred from differences in the levels of soil water within and outside the fairy circles, and from biomass measurements\cite{wesuls2010perceptions}.
The biomass growth parameter, $\Lambda$, was determined from estimates of the ratio of transpiration to dry biomass for tropical grass\cite{Erickson2012Water} and the previously estimated value of $\Gamma$.

Finally, the value of the soil-water diffusivity, $D_W$, was estimated using the Van Genuchten model for water retention curves \cite{vanGenuchten1980ClosedForm}.
Using parameters corresponding to sandy soil\cite{VGparams} and the values of soil-water content that correspond to those found in the field\cite{juergens2013biological}, we estimate $D_W$ to be in the range $0.1-100 [m^2/year]$.
We used a value that slightly exceeds the upper limit of this range to account for the possible enhancement of lateral water transport due to the network of underground channels formed by termite populations\cite{juergens2013biological}.
We emphasize that the model results reported here are robust and can be obtained with many different sets of parameters (see the SI Appendix (Sensitivity Analysis)).

\section{Earth Data Analysis}
We analyzed high resolution satellite images taken over the period of 2004-2013 in the NambiRand Nature Reserve.
The snapshots shown in Figs. 1 and 2 were processed from four images, captured in 2004, 2008, 2010 and 2013, whose spatial resolution is as high as $0.5 [m]$.
The rgb data shown were derived using Pan-Sharpening with ERDAS software when rgb was available (2004, 2010 and 2013), and false-color was used for the 2008 image.

The data used for Fig. 3 were taken from monthly precipitation data and from measurements of fairy circles features using satellite images.
The precipitation data were derived from interpolations of both the readings of meteorological stations within the NambiRand Nature Reserve (for the years 2000-2012)
and of the NOAA data center (for the years 1990-1999) for two locations in Namibia.
The numbers and sizes of the fairy circles were estimated from eight satellite images, the four previously described (from 2004, 2008, 2010 and 2013), and four others (from 2005, 2009, 2011 and 2012) taken from Google Earth data.
From each of the eight images, four regions of $180[m]x180[m]$ were chosen and analyzed.
A proprietary segmentation algorithm, involving morphological operations and thresholding with a CLAHE algorithm, was used to identify each fairy circle.
The identification results were then reviewed manually to minimize errors caused by the automatic process, and in total, about 800 fairy circles were identified and measured for each time point.

The cross-correlation measures shown in Fig. 3 were calculated by comparing the accumulated precipitation, $\psi$, with either the number of fairy circles, $\alpha$, or the average size of a fairy circle, $\beta$.
The accumulated precipitation is defined as:
$$ \psi_i\left(S\right) = \sum\limits_{j=0}^{S-1} P_{(i-j-L)}. $$
Here, $P_k$ is the precipitation at time $k$, $L$ is the lag time and $S$ is the integration time, with all times given in number of years.
In order to fit the rainy season in the field, we considered the annual precipitation between November 1 and the following October 31.
The lag time, $L$, was set to zero (i.e., considering the accumulated rain during a period $S$ ending at the end of the previous year) for the fairy circle size and $L=1$ for the fairy circle number; these values yield the most significant correlations between the precipitation and the fairy circle dynamics.
The cross-correlation for the fairy circle number $C_\alpha$ is defined as:
$$ C_\alpha\left(S\right) = \frac{ \sum\limits_{i=1}^N (\psi_i\left(S\right)-\overline{\psi\left(S\right)})\cdot(\alpha_i-\bar{\alpha}) }{ \sqrt{ \sum\limits_{i=1}^N (\psi_i\left(S\right)-\overline{\psi\left(S\right)})^2} \cdot  \sqrt{\sum\limits_{i=1}^N (\alpha_i-\bar{\alpha})^2}}.$$
$\bar{A}$ stands for the average of $A$ ($\bar{A}\equiv\frac{1}{N}\sum\limits_{i=1}^NA_i$), and $N$ is the number of time points, with the value of $8$ for the observation data, and either $8$ or $100$ for the simulation data.
The cross-correlation for the fairy circle size, $C_\beta\left(S\right)$, was calculated in the same manner.

\begin{acknowledgments}
We wish to thank Nils Odendaal, the Chief Executive Officer of the NamibRand Nature Reserve, for providing us the precipitation measurements of various meteorological stations within the reserve,
 and we also thank Elad Eizner and Noa Levi-Ohana for their help with the segmentation and analysis of the satellite data.
 The research leading to these results has received funding from the European Union Seventh Framework Programme (FP7/2007-2013) under grant number [293825], and from the Israel Science Foundation under grant number~305/13.
\end{acknowledgments}

\appendix
 
\section{Derivation of the simplified model}
We present the derivation of the simplified model that we use and analyze in the main text.
Our starting point is the vegetation model of Gilad et al. \cite{Gilad2007jtb} that includes three pattern-forming feedbacks.
Using the characteristics of the fairy circles ecosystem, we explain the assumptions and approximations that allow us to replace the integrals of the root-augmentation feedback with simpler algebraic terms and to decouple the aboveground water dynamics from the soil-water and vegetation dynamics.

The dryland vegetation model of Gilad et al. \cite{Gilad2007jtb} describes the coupled dynamics of the areal densities of vegetation biomass (\textit{B}), soil water (\textit{W}) and
surface water (\textit{H}), all having the dimension of mass per unit area. The dynamics is described with a temporal resolution that is smaller than the typical time scale for changes in the biomass (dictated by the biomass growth rate, mortality and dispersion rate) but is large enough to allow averaging the intermittent nature of the precipitation. The spatial scale resolution is larger than the scale of a single plant (in order to allow the description of a continuous vegetation density) but smaller than the typical patch size.  
Restricting our interest to flat terrains, the model reads:
\begin{subequations}\label{eq:ModelDM}
  \begin{align}
      B_T & = G_B B (1-B/K) - MB + D_B \nabla^2 B \\
      W_T & = IH - N(1-RB/K)W -G_W W + D_W \nabla^2 W \\
      H_T & = P - IH + D_H \nabla^2 (H^2)\,,
  \end{align}
\end{subequations}
where
\begin{subequations}
  \begin{align}
      G_B(\textbf{X},T) & = \Lambda \int_\Omega G(\textbf{X},\textbf{X}^\prime,T) W(\textbf{X}^\prime,T) \textbf{dX}^\prime \\
      G_W(\textbf{X},T) & = \Gamma \int_\Omega G(\textbf{X},\textbf{X}^\prime,T) B(\textbf{X}^\prime,T) \textbf{dX}^\prime \\
      G(\textbf{X},\textbf{X}^\prime,T) & = \frac{1}{2 \pi S_0^2} \exp \left[ -\frac{|\textbf{X}-\textbf{X}^\prime|^2}{2S_0^2(1+EB(\textbf{X},T))^2} \right] \label{eq:G} \\
      I & = A \frac{B(\textbf{X},T)+Qf}{B(\textbf{X},T)+Q}
  \end{align}
\end{subequations}

The terms $G_B$ and $G_W$ involve integration over a root kernel ($G$) that represents the spatial extent of the root zone in the lateral directions.
Applying the model to perennial grasses, e.g. \textit{Stipagrostis Ciliata} (common in the fairy circle ecosystem), which form patterns with a typical length scale of ~$10m$ \cite{picker2012ants} and have a root girth of approximately $0.5m$ \cite{Midgley1999Book}, allows us to assume that the kernel function is much narrower than the biomass and soil-water distributions. Under this condition, we can approximate the kernel by a Dirac delta function.
Formally, this is done by taking the limit $S_0 \rightarrow 0$ in Eq. \ref{eq:G}, where $S_0$ represents the lateral root-zone size of a seedling.
Following this approximation, we may replace the integrals of Eqs. \eqref{eq:G} by the algebraic forms:
\begin{subequations}
  \begin{align}
      G_B(\textbf{X},T) & = \Lambda W(\textbf{X},T) (1+EB(\textbf{X},T))^2;\\
      G_W(\textbf{X},T) & = \Gamma B(\textbf{X},T) (1+EB(\textbf{X},T))^2.
  \end{align}
\end{subequations}

The fairy circle ecosystem consists of sandy soil. This soil type is characterized by a high rate of surface water infiltration that is comparable to the infiltration rate in vegetated soil. To account for the absence of a significant infiltration contrast between bare and vegetated soil, we set $f=1$.
The infiltration rate then becomes a constant, $I=A$, independent of the biomass $B$, and the equation for the surface water variable $H$ decouples from those for $B$ and $W$.
This equation has a single stationary uniform solution, $H_0=P/I$, which is always linearly stable.
Since $H$ is the fastest variable, we can assume that, on the much slower time scales over which $B$ and $W$ significantly change, it has already equilibrated at $H_0$.
Inserting the solution $H=H_0$ into the equation for $W$, we obtain the two-variable model:
\begin{subequations}\label{eq:ModelDMVSG}
  \begin{align}
      B_T & = \Lambda W(\textbf{X},T) (1+EB(\textbf{X},T))^2 B (1-B/K) - MB + D_B \nabla^2 B \\
      W_T & = P - N(1-RB/K)W -\Gamma B(\textbf{X},T) (1+EB(\textbf{X},T))^2 W + D_W \nabla^2 W.
  \end{align}
\end{subequations}

\begin{figure}[!ht]
\includegraphics[width=0.95\linewidth]{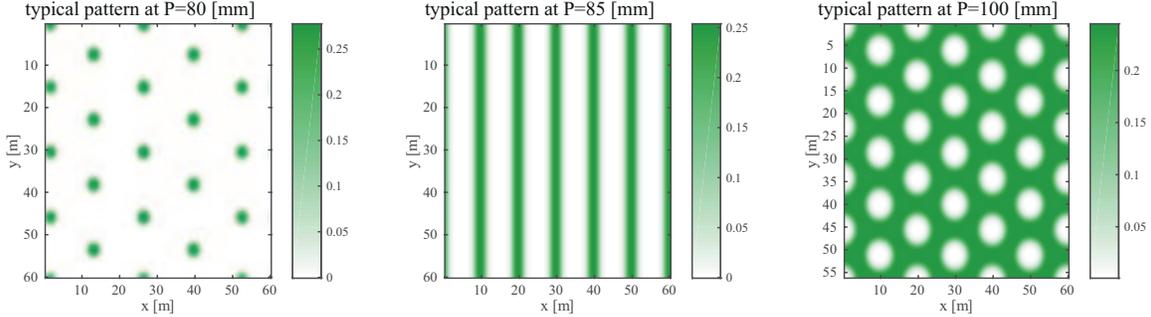}
 \caption{ Patterns predicted by the model for low precipitation rates.}
 \label{fig:patternslowp}
 \end{figure}

In the biomass equation, $\Lambda$ is the biomass growth rate coefficient,
$K$ is the maximal standing biomass,
$E$ is a measure for the root-to-shoot ratio, $M$ is the mortality rate, and $D_B$ represents the seed dispersal or clonal growth rate.
In the soil-water equation, $P$ is the precipitation rate, $N$ is the evaporation rate, $R$ is a dimensionless factor representing a reduction of the 
evaporation rate due to shading, $\Gamma$ is the water-uptake rate coefficient, and $D_W$ is the effective soil-water diffusivity in the lateral ($X,Y$) directions, 
assumed to be a constant, independent of the state variables, that represents linear diffusion\cite{Gilad2007jtb}. 
We refer the reader to an earlier publication\cite{Gilad2007jtb} for additional information about the original model.
The model considered here (Eqs. \eqref{eq:ModelDMVSG}) predicts, in addition to the states shown and discussed in the main paper, the appearance of spotted and patterned states under low precipitation rates as shown in Fig. \ref{fig:patternslowp}.
\section{Sensitivity Analysis}

In order to test the sensitivity of the results (presented in the main text) to the set of parameters, we first write the model in a dimensionless form.
Thereby, we reduce the number of parameters and simplify the analysis.

\begin{table}[h]
  \begin{tabular}{ | c || c | c| }
    \hline
    parameter & \ \ \ \ $+10\%$ \ \ \ \ & \ \ \ \ $-10\%$ \ \ \ \ \\ \hline \hline
    $\lambda$ & $0.8913 $ & $1.1269 $ \\ \hline 
    $\eta$ & $1.0179 $ & $0.9664 $ \\ \hline 
    $\nu$ & $1.0604 $ & $0.9404 $ \\ \hline 
    $\rho$ & $1.0183 $ & $0.9819 $ \\ \hline 
    $\delta_w$ & $1.0556 $ & $0.9413 $ \\ \hline 
     \end{tabular}
       \caption{ Relative change in the existence range of the hybrid states (snaking range), due to varied parameters compared to those used in the main text. 
  In each case, one parameter is changed by either $+10\%$ or $-10\%$, and the relative existence range is shown.}
\end{table}

\begin{figure}[!ht]
\includegraphics[width=0.95\linewidth]{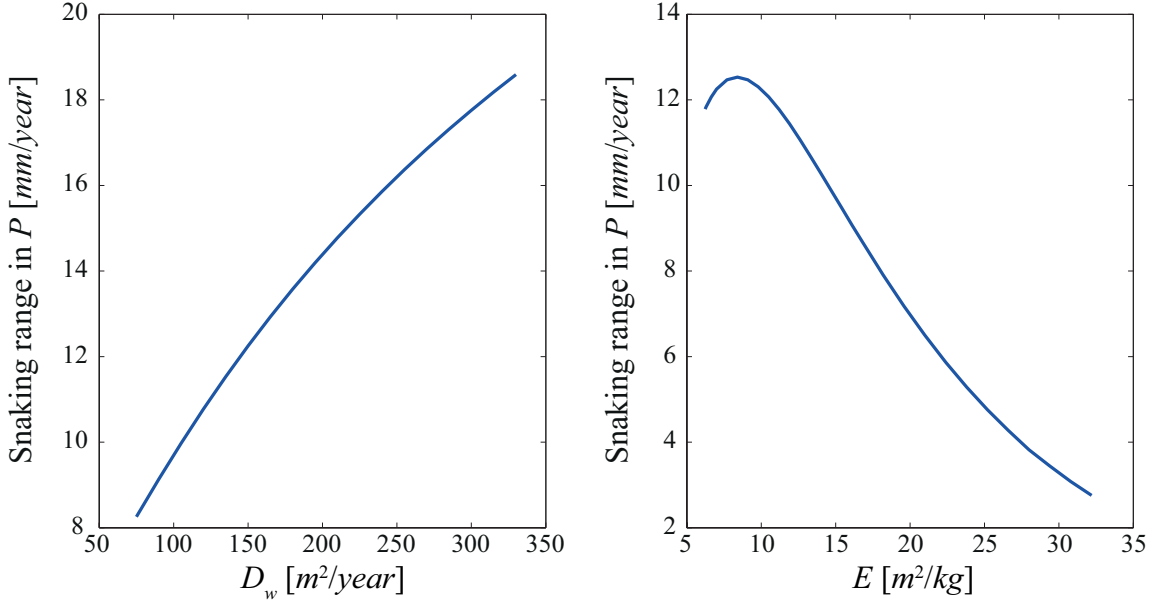}
 \caption{ The range of precipitation rates, $P$, for which stable hybrid states exist, as a function of either $D_W$ (left) or $E$ (right).}
 \label{fig:ERedw}
 \end{figure}

\begin{figure}[!ht]
\includegraphics[width=0.95\linewidth]{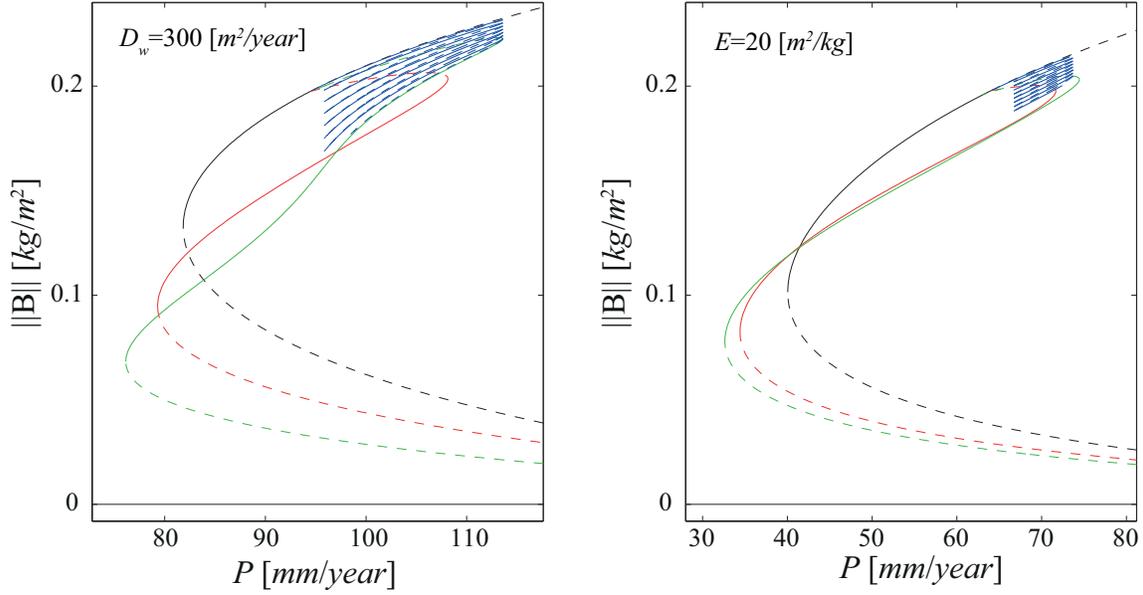}
 \caption{Bifurication graphs for two sets of parameters. 
 On the left panel, $D_W$ was changed to $300$ [$m^2$/$year$], while on the right panel, $E$ was changed to $20$ [$m^2$/$kg$].
 Uniform states (bare soil and uniform vegetation) are shown in black, two periodic states are shown in red and green, and the hybrid states are shown in blue.
 }
 \label{fig:BifExamples}
 \end{figure}

Equations \eqref{eq:ModelDMVSG} describe the dynamics of the aboveground biomass ($B$) and the soil water ($W$) in a dimensional form. 
Translation of the model equations to a dimensionless form is achieved by rescaling the state variables $B,W$ and the space and time coordinates as follows:
\begin{align}
    b = \frac{B}{K}; w = \frac{W\Lambda}{K \Gamma}; t = M T; x = X\sqrt{M/D_B}\,.
\end{align}
In terms of these dimensionless quantities, the model reads:
\begin{equation}
\partial_t b = \lambda w b(1 + \eta b)^2(1 - b) - b + \nabla^2b \,, \label{ndSGeqb} \\
\end{equation}
\begin{equation}
\partial_t w = p - \nu w (1-\rho b) - \lambda w b(1 + \eta b)^2  + \delta_w\nabla^2 w\,. \label{ndSGeqw}  \\
\end{equation}
The dimensionless parameters here are related to their dimensional counterparts by the following relations:
\begin{align}
  &  \lambda = \frac{K \Gamma}{M}; \ \ \ \ \eta = E K;  \ \ \ \   p = \frac{\Lambda P}{K \Gamma M} ; \ \ \ \   \nu = \frac{N}{M};  \ \ \ \   \rho = R ;   \ \ \ \  \delta_w = \frac{D_W}{D_B} \,.
\end{align}

We are thus left with only six parameters, for which we use $p$ as the main bifurcation parameter, and as such, its value changes throughout our analysis.
The basis for the main results is the existence of stable hybrid states; therefore, we tested the conditions under which they occur and their existence range (range of precipitation rate, $p$).
We changed the other five parameters, one at a time, within $\pm 10\%$ of their value. We found that in all cases, the hybrid states exist, with their existence range slightly changing.
The relative changes in their existence range are shown in Table I.

As can be seen, the parameter $\lambda$ has the strongest effect on the snaking range.
The parameters $\eta$ and $\delta_w$ control the pattern-forming feedback in our model \cite{Kinast2014prl}. 
Therefore, we present a more thorough analysis of their effect on the snaking range. 
In Fig. \ref{fig:ERedw}, we show the snaking range as a function of either $D_W$ (left panel) or $E$ (right panel). 
Fig. \ref{fig:BifExamples} shows two examples of the bifurcation diagram for a relatively big change in one of these parameters.

\begin{figure}[!ht]
\includegraphics[width=0.55\linewidth]{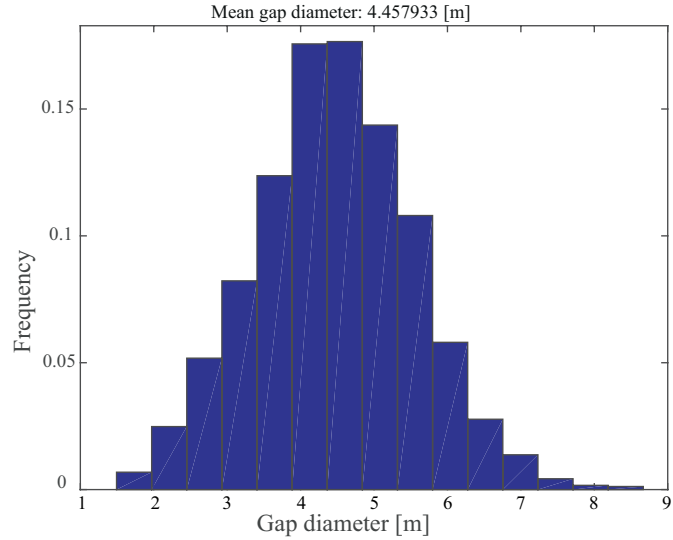}
 \caption{ Histogram of the FCs size in the regions covered by the satellite images that were analyzed.}
 \label{fig:FCsizehist}
 \end{figure}
%

In addition to the sensitivity of the snaking range to the parameters the size of the gaps and the distance between neighboring gaps are also affected by the parameters.
The histogram of the FC sizes, in the regions for which the satellite images were analyzed, is shown in Fig. \ref{fig:FCsizehist}.
\begin{figure}[!ht]
\includegraphics[width=0.95\linewidth]{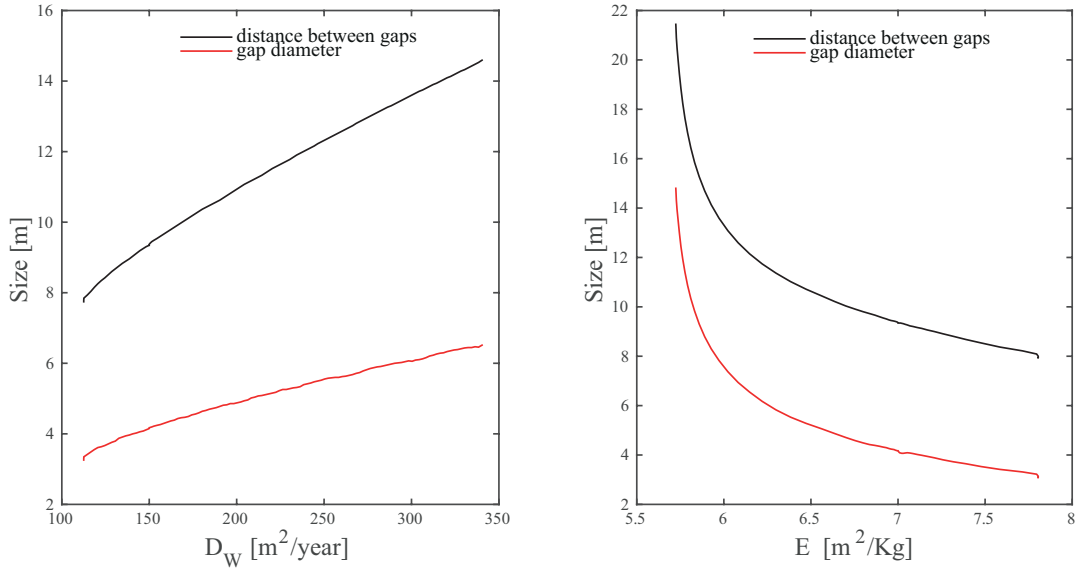}
 \caption{ The average gap size and distance between neighboring gaps vs. the parameters $D_w$ (left panel) and $E$ (right panel) as 
 predicted by the model for constant $P=102 [mm/yr]$.}
 \label{fig:Modelsize}
 \end{figure}
In Fig. \ref{fig:Modelsize} we show the average size of the FCs and the average distance between them as predicted by our model for constant precipitation rate of $P=102 [mm/yr]$ and for a range of the
feedback control parameters $D_w$ and $E$. It is shown that there is a range of values of these parameters that correspond to the FCs size observed in the field.
The range of possible values for these parameters, which have not been measured directly in the field, may be estimated from these graphs.    

\section{Predicted response of uniform vegetation to a series of droughts}

\begin{figure}[!ht]
\includegraphics[width=0.95\linewidth]{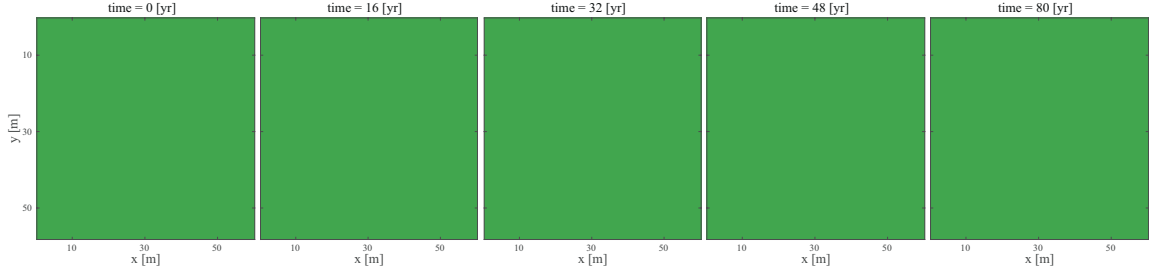}
 \caption{ The response of 2D uniform vegetation to periodic short droughts. No regime shift takes place even though the system is subjected to the same rainfall regime as the 
 hybrid state in figure 4\textbf{B}. The rainfall regime represents periodic one-year droughts with $P=81 [mm/yr]$, followed by 15 years of $P=102 [mm/yr]$.
 The simulated domains are $60 [m]\times56 [m]$.}
 \label{fig:UnResponse}
 \end{figure}

In the main paper, we showed that the response of vegetation to a series of droughts or spates depends on the initial condition.
To complement the information we present in Fig. \ref{fig:UnResponse} the response of a uniform vegetation to a series of droughts.


\end{document}